\documentclass[prd, twocolumn, lengthcheck, superscriptaddress, showpacs, letterpaper, nofootinbib]{revtex4-1}
\usepackage{hyperref}
\usepackage{times}
\usepackage{amsmath}
\usepackage{graphicx}
\usepackage{acronym}
\usepackage{color}
\usepackage{microtype}

\newcommand\given{\,|\,}

\newcommand\sig{\mathbf{h}}
\newcommand\data{\vec{c}}

\newcommand\Ft{\ensuremath{F^\ast}}
\newcommand\Pt{\ensuremath{P^\ast}}
\newcommand\ct{\vec{\ensuremath{c^\ast}}}

\newcommand\diff{\, \mathrm{d}}

\newcommand\mc{\mathcal{M}}
\def\Msun{\ensuremath{M_{\odot}}}

\def\thercsid{\relax}
\def\rcsid#1{\def\next##1#1{\def\thercsid{##1}}\next}
\rcsid$Id: likelihood.tex,v 1.77 2012/01/11 17:24:02 vaulin Exp $

\begin{document}
\title{Likelihood-ratio ranking of gravitational-wave candidates in a non-Gaussian
background.}
%\title{Detection of gravitational waves in the presence of non-Gaussian background noise}

\author{Rahul Biswas}
\affiliation{University of Texas-Brownsville, Brownsville, Texas 78520, USA}

\author{Patrick R. Brady}
\affiliation{University of Wisconsin--Milwaukee, Milwaukee, WI 53201, USA}

\author{Jordi Burguet-Castell}
\affiliation{Universitat de les Illes Balears, E-07122 Palma de Mallorca, Spain}

\author{Kipp Cannon}
\affiliation{Canadian Institute for Theoretical Astrophysics, University of Toronto, Toronto, Ontario, M5S 3H8, Canada}

\author{Jessica Clayton}
\affiliation{University of Wisconsin--Milwaukee, Milwaukee, WI 53201, USA}

\author{Alexander Dietz}
\affiliation{The  University of Mississippi, University, MS 38677, USA}

\author{Nickolas Fotopoulos}
\affiliation{LIGO - California Institute of Technology, Pasadena, CA 91125, USA}

\author{Lisa M. Goggin}
\affiliation{University of California San Francisco, San Francisco, CA 94143 USA}

\author{Drew Keppel}
\affiliation{Albert-Einstein-Institut, Max-Planck-Institut f\"ur Gravitationsphysik, D-30167 Hannover, Germany}
\affiliation{Leibniz Universit\"at Hannover, D-30167 Hannover, Germany}

\author{Chris Pankow}
\affiliation{University of Wisconsin--Milwaukee, Milwaukee, WI 53201, USA}

\author{Larry R. Price}
\affiliation{LIGO - California Institute of Technology, Pasadena, CA 91125, USA}

\author{Ruslan Vaulin}
\affiliation{LIGO - Massachusetts Institute of Technology, Cambridge, MA 02139, USA}

\date{\thercsid}

%DCC number  LIGO-P1100196

\begin{abstract}
We describe a general approach to detection of transient
gravitational-wave signals in the presence of non-Gaussian background noise. We
prove that under quite general conditions, the ratio of the likelihood of
observed data to contain a signal to the likelihood of it being a noise
fluctuation provides optimal ranking for the candidate events found in an
experiment. The likelihood-ratio ranking allows us to combine different kinds of data into a single analysis. We apply the general framework to the problem of unifying the
results of independent experiments and the problem of accounting for
non-Gaussian artifacts in the searches for gravitational waves from compact
binary coalescence in LIGO data. We show analytically and confirm through
simulations that in both cases the likelihood ratio statistic results in an
improved analysis.
\end{abstract}

\maketitle
%acronyms
\acrodef{SNR}{signal-to-noise ratio}
\acrodef{LIGO}{Laser Interferometer Gravitational-wave Observatory}

%%%%%%%%%%%%%%%%%%%%%%%%%%%%%%%%%%%%%%%%%%%%%%%%%%%%%%%%%%%%%%%%%%%%%%%%
\section{Introduction}
\label{sec:intro}
%%%%%%%%%%%%%%%%%%%%%%%%%%%%%%%%%%%%%%%%%%%%%%%%%%%%%%%%%%%%%%%%%%%%%%%%

The detection of gravitational waves from astrophysical sources is a
long-standing problem in physics. Over the past decade, the experimental emphasis
has been on the construction and operation of kilometer-scale interferometric detectors such as \ac{LIGO}~\cite{Abbott:2007kv}. The instruments measure the
strain, $s(t)$, by monitoring light at the interferometer's output port, which varies as test
masses that are suspended in vacuum at the ends of orthogonal arms differentially
approach and recede by minuscule amounts. The strain signal, $s(t)$, is
a combination of noise, $n(t)$, and gravitational-wave signal, $h(t)$.

There is a well established literature describing the analysis of time-series
data for signals of various types~\cite{wainstein:1962}; these methods have been
extended to address gravitational-wave detection~\cite{FinnChernoff:1993}. This
approach usually begins with the assumption that the detector noise, $n(t)$, is
stationary and Gaussian. Then one proceeds to derive a set of filters that are
tuned to detect the particular signals in this time-series data. The result is
both elegant and powerful: whitened detector noise is correlated with a
whitened version of the expected signal. The approach has been used to develop
techniques to search for gravitational waves from compact binary coalescence, isolated
neutron stars, stochastic sources, and generic bursts with certain
time-frequency characteristics~\cite{Anderson:2000yy}.

This approach takes the important first step of designing filters that
properly suppress the dominant, frequency-dependent noise sources in the
instrument. The simplicity of the filters is due to the fact that the power-spectral density fully characterizes the statistical properties of stationary, Gaussian noise. However, interferometric detectors are prone to non-Gaussian and
non-stationary noise sources. Environmental disturbances, including seismic, acoustic, and
electromagnetic effects, can lead to artifacts in the time series that are
neither gravitational waves nor stationary, Gaussian noise. Imperfections in
hardware can lead to unwanted signals in the time series that originate from auxiliary
control systems. 

To help identify and remove these unwanted signals, instruments have been
constructed at geographically separated sites and the data are analyzed
together. A plethora of diagnostics have also been developed to characterize
the quality of the data~\cite{Christensen:2010,Robinet:2010zz,SeisVeto}. Searches for gravitational waves use
more than just the filtered output of the time-series, $s(t)$, to
separate gravitational-wave signals from noise. Moreover, the responses from
various filters indicate that the underlying noise sources are not Gaussian,
even after substantial data quality filtering and coincidence requirements have
been applied. 

In this paper, we discuss using likelihood-ratio ranking as a unified approach
to gravitational-wave data analysis. The approach foregoes the stationary,
Gaussian model of the detector noise. The output of the filters derived under
that assumption becomes one element in a list of parameters that characterize a
gravitational-wave detection candidate. The detection problem is then couched
in terms of the statistical properties of an $n$-tuple of derived quantities,
leading directly to a likelihood-ratio ranking for detection candidates. The
$n$-tuple can include more information than simply the \ac{SNR}
measured in each instrument of the network. It can include measures of data
quality, the physical parameters of the gravitational-wave candidate, the
\ac{SNR} from the coherent and null combinations of the detector
signals; it can include nearly any measure of detector behavior or signal
quality.  

This approach was already used to develop ranking statistics for compact binary
coalescence signals~\cite{GRB070201, S5GRB, S5LowMassLV} and is at the core of
a powerful coincidence test developed for burst
searches~\cite{0264-9381-25-10-105024}.

This work presents a general framework for the likelihood-ratio ranking in the context of gravitational-wave detection. We explore its analytical properties and illustrate its practical value by applying it to two data analysis problems arising in real-life searches for gravitational waves in LIGO data.

\section{General derivation of likelihood-ratio ranking}
\label{GeneralDerivation}

Let the $n$-tuple $\data$ denote the observable data in some experiment that aims to detect a signal denoted by $\sig$. This signal can usually
be parametrized by several continuous parameters that may be unknown, for example
distance to the source of gravitational waves and location on the sky. The
purpose of the experiment is to identify the signal. Depending on whether a Bayesian or frequentist statistical approach is taken, this is stated in terms of either the probability that the signal is present or the probability
that the observed data are a noise fluctuation.

In this section, we show that both approaches lead to ranking candidate signals
according to the likelihood ratio
\begin{equation}
\label{eq:likelihoodratio}
\Lambda(\data) = \frac{\int \! p(\data \given \sig, 1) p(\sig \given 1) \diff \sig}
                      {p(\data \given 0)}\,,
\end{equation}
where $p(\data \given \sig, 1)$ is the probability of observing $\data$ in the
presence of the signal $\sig$, $p(\sig \given 1)$ is the prior probability to
receive that signal, and $p(\data \given 0)$ is the probability of observing
$\data$ in the absence of any signal. The higher a candidate's $\Lambda$ value, the more likely it is a real signal.

\subsection{Bayesian Analysis}
\label{Bayesian Analysis}

In this approach, we compute the probability that a signal is present in the
observed data, $p(1 \given \data)$. By a straightforward application of Bayes
theorem, we write
\begin{align}
\label{bayes_prob}
p(1 \given \data) & =
  \frac{p(\data \given 1) p(1)}
       {p(\data \given 1) p(1) + p(\data \given 0) p(0)} \nonumber\\
         & =
  \frac{\int \! p(\data \given \sig, 1) p(\sig \given 1) p(1) \diff \sig}
       {\int \! p(\data \given \sig, 1) p(\sig \given 1) p(1) \diff \sig + p(\data \given 0) p(0)}\,,
\end{align}
where $p(0)$ is the \emph{a priori} probability that the signal is absent and
$p(1)$ is the \emph{a priori} probability that there is a signal (of any kind). The
denominator re-expresses $p(\data)$ in terms of the two possible independent
outcomes: the signal is present or the signal is absent. Upon successive
division of numerator and denominator by $p(\data \given 0)$ and $p(1)$, we
find
\begin{equation}
\label{probsiggivendata}
p(1 \given \data) = \frac{\Lambda(\data)}{\Lambda(\data) + p(0) / p(1)}\,,
\end{equation}
which is a monotonically increasing function of the likelihood ratio $\Lambda$
defined by Eq.~\eqref{eq:likelihoodratio}~\footnote{This ratio of likelihoods
is also known as the \emph{Bayes factor}.}. Hence, the larger the likelihood
ratio, the more probable it is that a signal is present.

\subsection{Frequentist Approach}
\label{FrequentistApproach}

The process of detection can always be reduced to a binary ``yes'' or ``no'' question---does the observed data contain
the signal? An optimal detection scheme should achieve the maximum rate of
successful detections---correctly given ``yes'' answers---with some fixed,
preferably low, rate of false alarms or false positives---incorrectly given
``yes'' answers. This is the essence of the Neyman-Pearson optimality criteria
for detection, which states that an optimal detector should maximize the
probability of detection at a fixed probability of false
alarm~\cite{neyman-1933}.

As before, let the $n$-tuple $\data$ denote the observable data and $\sig$ the
signal that is the object of the search. Without loss of generality, any
decision-making algorithm can be mapped into a real function, $f(\data)$, of
the data that signifies detection whenever its value is greater than or equal
to a threshold value, $\Ft$. Thus, using the Neyman-Pearson formalism, an optimal detector is realized by finding a function,
$f(\data)$, that maximizes the probability of detection at a fixed value of
the probability of false alarm. The probability of detection, $P_{1}$, is%
\begin{equation}
\label{detprob}
P_{1} = \int_{V_{d}} \int_{V_{\sig}} \! \Theta\left(f(\data) - \Ft \right)p(\data \given \sig, 1) p(\sig \given 1) p(1) \diff \sig \diff \data\,, 
\end{equation}
and the probability of false alarm~\footnote{This is similar, but not exactly equal, to the \emph{false-alarm probability} or \emph{Type I error}, which assumes the case where no signal is present, that is, does not include the term $p(0)$.}, $P_{0}$, is
\begin{equation}
\label{faprob}
P_{0} = \int_{V_\mathrm{d}} \! \Theta\left(f(\data) - \Ft \right)p(\data \given 0) p(0) \diff \data \,,
\end{equation}
where $V_{\sig}$ identifies the subset of signals targeted by the search, $V_\mathrm{d}$ denotes the subset of accessible data and integration is performed over all signals, $\sig$, and data points, $\data$, within these subsets. Treating $P_{1}$ and $P_{0}$ as functionals of $f(\data)$, we find that for an
optimal detector, the variation of
\begin{equation}
\label{Sfunc}
S[f(\data)] = P_{1}[f(\data)] - l_0\left( P_{0}[f(\data)] - \Pt \right)
\end{equation}
should vanish. Here $l_0$ denotes a Lagrange multiplier and \Pt is a constant
that sets the value of the probability of false alarm. The variation of Eq.~\eqref{Sfunc} with respect to $f(\data)$ gives
\begin{align}
\delta S =& \int_{V_\mathrm{d}} \! \delta\left(f(\data) - \Ft \right)\delta f(\data) \nonumber\\
 & \times \left[ \int_{V_{\sig}} p(\data \given \sig, 1) p(\sig \given 1) p(1) \diff \sig - l_0 p(\data \given 0) p(0) \right]\diff \data \,.
\end{align}
Variations $\delta f(\data)$ at different data points are independent, thus
implying that after integration over $\data$, the condition
\begin{equation}
\label{optcondition}
\frac{\int \! p(\ct \given \sig, 1) p(\sig \given 1) \diff \sig}{p(\ct \given 0)} = \frac{l_0 p(0)}{p(1)} = \text{const}
\end{equation}
must be satisfied at all points $\ct$ for which the argument of the delta
function satisfies the condition
\begin{equation}
f(\ct) - \Ft = 0 \,.
\end{equation}
This latter condition defines the detection surface separating the detection and non-detection regions. Note that as
$f(\data)$ varies, the shape of this surface changes accordingly. Therefore,
Eq.~\eqref{optcondition} implies that the detection surface
must be the surface of the constant likelihood ratio for an optimal detector. This is the only
condition on the functional form of $f(\data)$. Variation with respect to $\Ft$
does not give a new condition, whereas variation with respect to the Lagrange
multiplier, $l_0$, simply sets the probability of a false alarm to be $\Pt$.~\footnote{In the case of the mixed data, when $\data$ includes continuous as well as discrete parameters, integration in the expressions for $P_1$ and $P_0$ should be replaced by summation wherever it is appropriate. This does not affect the derivation or the main result. The notion of optimal detection surface defined by Eq.~(\ref{optcondition}) is straightforward to generalize to include both continuous and discrete data.}

A natural way to satisfy the optimality criteria is to use the likelihood ratio
\begin{equation}
\label{likelihooddef}
\Lambda(\data) = \frac{\int \! p(\data \given \sig, 1) p(\sig \given 1) \diff \sig}{p(\data \given 0)}
\end{equation}
or any function $f(\Lambda(\data))$ for ranking the candidate signals. With
this choice, the optimality condition Eq.~\eqref{optcondition} is satisfied
for any threshold $\Ft$. The latter is determined by the choice of an
admissible value of the probability of false alarm, $\Pt$, through
\begin{equation}
 P_{0}[f(\Lambda(\data))] = \Pt \,.
\end{equation}

\subsection{Variation of efficiency with volume of search space}
\label{appendix:effproof}

The likelihood ratio given in Eq.~\eqref{eq:likelihoodratio} is guaranteed to maximize
the probability of signal detection for a given search. Because
optimization is performed for a fixed region defined by all possible values of a candidate's parameters, $\data$, in the search, it is unclear whether
increasing the volume of available data (e.g. extension of the bank of template
waveforms) would not result in an overall decrease of probability to detect
signals. For example, one may be apprehensive of the potential increase in the
rate of false alarms solely due to extension of the parameter space.
Intuitively, having more available information should not negatively
affect the detection probability or efficiency if the information is processed
correctly.  In what follows, we prove that this is true if the
likelihood ratio is used for making the detection decision.

To prove that the detection efficiency does not decrease when the
volume of data is increased, we must show that the variation of $\delta P_1/
\delta V_\mathrm{d}$ at a fixed $P_0$ is non-negative. Consider a foliation of the space of data, $V_\mathrm{d}$, by surfaces of
constant likelihood ratio, $S_\Lambda$. Functionals for the probabilities of
detection, Eq.~\eqref{detprob}, and false alarm, Eq.~\eqref{faprob} can be written as
\begin{equation}
P_1 = \int_0^\infty \! \diff \Lambda \int_{S_\Lambda} \! \Theta\left(\Lambda - \Lambda^*\right) p(\data\given 1) p(1) \,,
\end{equation}
and
\begin{equation}
P_0 = \int_0^\infty \! \diff \Lambda \int_{S_\Lambda} \! \Theta\left(\Lambda - \Lambda^*\right) p(\data\given 0) p(0)\,,
\end{equation}
where, for brevity, we absorbed explicit integration in the space of signals, $V_{\sig}$, in the product $ p(\data\given 1) p(1)$. $P_1$ is a functional of $V_\mathrm{d}$ and $\Lambda^*$. Since the latter is
determined by the value chosen for false alarm probability, $P_0 = P^*$, and
the probability of false alarm also depends on $V_\mathrm{d}$,
variations of $V_\mathrm{d}$ and $\Lambda^*$ are not independent. To find the
relation, we vary the probability of false alarm
\begin{equation}
\begin{split}
\delta P_0 =& - \int_0^\infty \! \diff \Lambda \int_{S_\Lambda} \! \delta \left(\Lambda - \Lambda^*\right) p(\data\given 0) p(0) \delta \Lambda^* \\
& + \int_0^\infty \! \diff \Lambda \int_{\delta S_\Lambda} \! \Theta \left(\Lambda - \Lambda^*\right) p(\data\given 0) p(0) \delta S_\Lambda\,.
\end{split}
\end{equation}
We consider non-negative variations of surfaces of constant
likelihood ratio, $\delta S_\Lambda$, that correspond only to the addition of
new data points to $V_\mathrm{d}$, and therefore correspond only to an
extension of surfaces, $S_\Lambda$, without an overall translation or change of
shape.

The probability of false alarm should stay constant, therefore its variation
should vanish, providing the relation
\begin{equation}
\label{LambdaVdrelation}
\delta \Lambda^* = \frac{\int_0^\infty \! \diff \Lambda \int_{\delta S_\Lambda} \! \Theta \left(\Lambda - \Lambda^*\right) p(\data\given 0) \delta S_\Lambda } {\int_{S_{\Lambda^*}} \! p(\data \given 0)} \,.
\end{equation}
Next, we vary the functional for the detection probability
\begin{equation}
\begin{split}
\delta P_1 =& - \Lambda^* \int_{S_{\Lambda^*}} \! p(\data \given 0) p(1) \delta \Lambda^* \\
& + \int_0^\infty \! \diff \Lambda \int_{\delta S_\Lambda} \! \Theta \left(\Lambda - \Lambda^*\right) \Lambda \, p(\data\given 0) p(1) \delta S_\Lambda \,,
\end{split}
\end{equation}
where we use $p(\data \given 1) = \Lambda(\data)p(\data \given 0)$, which
follows from the definition of the likelihood ratio. Eliminating $\delta \Lambda^*$
by means of Eq.~\eqref{LambdaVdrelation} and re-arranging terms we get
\begin{equation}
\delta P_1 = p(1)\int_0^\infty \! \diff \Lambda \int_{\delta S_\Lambda} \! \Theta \left(\Lambda - \Lambda^*\right) \left(\Lambda - \Lambda^*\right) p(\data \given 0) \delta S_\Lambda \,,
\end{equation}
which is non-negative for all positive $\delta S_\Lambda$ by virtue of $\Theta
\left(\Lambda - \Lambda^*\right) \left(\Lambda - \Lambda^*\right)
\geq 0$. This proves that if the likelihood-ratio statistic is used in the detection process, the probability of detection can never decrease during an extension of the volume of available data.

\section{Applications}

In Section~\ref{s:combining_disjoint_experiments}, we apply the formalism of Section~\ref{GeneralDerivation} when assessing the significance of triggers between experiments on disjoint times. In Section~\ref{s:combining_search_spaces}, we demonstrate how the likelihood-ratio ranking can improve analysis efficiency by accounting for non-Gaussian features in the distributions of parameters of the candidates events.

\subsection{Combining disjoint experiments}
\label{s:combining_disjoint_experiments}

One complexity that arises in real-world applications is the necessity to
combine results from multiple independent experiments. For example, gravitational-wave searches are often thought of in terms of times when a fixed number of
interferometers are operating. If a network consists of
instruments that are not identical and located at different places, each
combination of operating interferometers may have very different combined
sensitivity and background noise. Times when three interferometers are
recording data may be treated differently from those when any pair is
operating. Ideally, these experiments would be treated together
accounting for differences in detectors' sensitivities and background noise in
the ranking of the candidate signals, but it is often not practical (see
how this problem was addressed in \cite{S5LowMassLV}). In this section, we show
that the likelihood-ratio ranking offers a natural solution to this problem, which is
conceptually similar to a simplified approach taken in~\cite{S5LowMassLV}.

Consider a situation in which the data is written as $\data = ( \vec{d}, j)$,
where $j=0,1,2,\ldots$ indicates that the data arose from an experiment
covering some time interval $T_j$. Note that $T_i \cap T_j = \emptyset$ if
$j\neq i$. The probability that a signal is present given the data is
\begin{equation}
\label{psignalforcombexp}
p(1 \given \vec{d},j) = \frac{\int \! p(\vec{d},j \given \sig, 1) p(\sig \given 1)p(1) \diff \sig}{
\int \! p(\vec{d},j \given \sig, 1 ) p(\sig \given 1)p(1) \diff \sig + p(\vec{d},j \given 0) p(0)} \,.
\end{equation}
 The conditional probabilities for the observed data can be further expanded as
\begin{align}
p(\vec{d}, j\given \sig, 1)& = p(\vec{d} \given j, \sig, 1)p(j \given \sig, 1)\\
p(\vec{d}, j \given 0)& = p(\vec{d} \given j, 0)p(j \given 0)\,,
\end{align}
where we introduce $p(j \given \sig, 1)$ and $p(j \given 0)$---the probabilities
for the observed data to be from the $j^\mathrm{th}$ experiment given the presence or
absence of a signal respectively. It is reasonable to assume\footnote{This is not strict equality. Gravitational-wave events can alter the amount of live time in experiments to detect them. For example, an alert sounds in the LIGO and Virgo control rooms when gamma-ray bursts are detected, which sometimes accompany CBCs. The alert prompts operators to avoid routine maintenance and hardware injections, with their associated deadtimes, for the following forty minutes.} that $p(j \given \sig, 1) = p(j\given 0)$. In this case, both probabilities drop out of Eq.~\eqref{psignalforcombexp}, and the expression for the probability for a signal to be present in the data can be written as 
\begin{equation}
\label{psignalforcombexpwithlambda}
p(1 \given \vec{d}, j) = \frac{\Lambda(\vec{d}, j)}{\Lambda(\vec{d}, j) + p(0) / p(1)}\,,
\end{equation}
with the likelihood ratio $\Lambda(\vec{d}, j)$ given by
\begin{equation}
\label{timecatlikelihood}
\Lambda(\vec{d}, j) = \frac{\int \! p(\vec{d} \given j, \sig, 1)p(\sig) \diff \sig}{p(\vec{d} \given j,0)}\,.
\end{equation}

Comparing Eq.~\eqref{psignalforcombexpwithlambda} with
Eq.~\eqref{probsiggivendata}, we conclude that the likelihood ratio
$\Lambda(\vec{d}, j)$, evaluated independently for each experiment, provides
optimal unified ranking. In terms of their likelihood ratios, data samples
from different experiments can be compared directly, with differences in
experiments' sensitivities and noise levels being accounted for by $p(\vec{d}
\given j, \sig, 1)$ and $p(\vec{d} \given j,0)$. 
 
Following the steps outlined in Section~\ref{FrequentistApproach}, the same
result can also be attained by direct optimization of the combined probability of detection at the fixed probability of false alarm. Optimality guarantees that the results of the less sensitive experiment can be combined with the results of the more sensitive experiment without loss of efficiency. In this approach, a unified scale provided by the likelihood ratio, $\Lambda(\vec{d}, j)$, is explicit because, by construction, the same threshold, $\Ft$, is applied to all data samples. $\Ft$ is determined by the value of the probability of false alarm for the combined experiment, given by
\begin{equation}
\label{timecatfaprob}
P_{0} = \sum_j \int \! \Theta\left(\Lambda(\vec{d}, j) - \Ft \right)p(\vec{d} \given j, 0)p(j \given 0) p(0) \diff \vec{d}\,,
\end{equation}
which makes the whole process less trivial. Notice that $p(j \given 0)$ (often approximated by $T_j /\sum_i T_i$) appears in the expression for $P_{0}$, however it does not appear in the expression for the likelihood ratio given in Eq.~\eqref{timecatlikelihood}. Since $p(j \given 0)$ is
proportional to the experiment duration, $T_j$, each experiment is weighted appropriately in the total probability of false alarm. In
a similar fashion, experiment durations appear in the expression for the
combined efficiency or the probability of detection for the combined
experiment.  
 
%Summarizing the results of this section, we found that the optimal detection strategy in the case of disjoint epochs is (i) rank candidate events using likelihood ratio (\ref{timecatlikelihood}) calculated for each epoch (ii) evaluate false-alarm probability (\ref{timecatfaprob}) as a function of $\Ft$ for all epochs combined (iii) find detection threshold $\Ft$ by choosing appropriate value for probability of false alarm (iv) from each epoch accept only those candidates whose likelihood ratio is greater or equal to the detection threshold $\Ft$.

\subsection{Combining search spaces}
\label{s:combining_search_spaces}

Sophisticated searches for gravitational-wave signals from compact binary
coalescence~\cite{LIGOS3S4all, Collaboration:2009tt,
Abbott:2009qj, S5LowMassLV} have been developed over the past decade. The non-Gaussian and
non-stationary noise is substantially suppressed by the application of
instrumental and environmental vetoes~\cite{Christensen:2010,Robinet:2010zz,SeisVeto}, coincidence between detectors, and
numerous other checks on the quality of putative gravitational-wave signals.
Nevertheless, the number of background triggers as a function of \ac{SNR}
depends on the masses of the binaries targeted in a search. For this reason,
triggers have been divided into categories based on the chirp mass, $\mc$, of the filters that produced the trigger (where $\mc=((m_1m_2)^3/(m_1+m_2))^{1/5}$ and $m_1$ and $ m_2$ are the masses of the compact objects in the binary). The background is a slowly varying function of $\mc$, falling off more rapidly, as a function of \ac{SNR}, for smaller values of $\mc$. This is a manifestation of non-Gaussianities still present in the data. It is desirable to account for this dependence when ranking candidates found in the search.

In this section, we consider a toy problem that mimics the properties of the
compact binary search but demonstrates how the likelihood-ratio ranking matches our
intuition. Following that example, we present the results of a simulated
compact binary search and demonstrate that the detection statistic based on
the likelihood ratio accounts for non-Gaussian features in background distribution and improves search efficiency. 

\subsubsection{Toy Problem}

Consider an experiment in which the data that define a candidate are $\data =
(\rho, x)$, where $\rho$ is the \ac{SNR} and $x$ is the extra parameter
describing the data sample (e.g. the chirp mass of the binary). Suppose the
distribution of the data in the absence of a signal is
%
%\graphicspath{{likelihood_plots/}}
\begin{figure}
\includegraphics[width=3in]{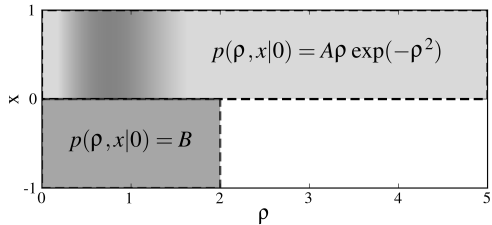}
\caption{Graphic representation of the model background distribution of
Eq.~\eqref{toy_distribution} for $\alpha = 2.0$. Shaded areas define the
regions of non-zero probability. }
\label{fig:toy_problem}
\end{figure}
\begin{equation}
\label{toy_distribution}
\begin{split}
p(\rho, x \given 0) =& A \rho \exp(-\rho^2) \Theta(x) \Theta(1-x)\\
& + B \Theta(x+1) \Theta(-x) \Theta(\rho) \Theta(\alpha-\rho) \,.
\end{split}
\end{equation}
Figure~\ref{fig:toy_problem} provides a graphic representation of this
distribution. Notice that $p(\rho,x \given 0)=0$ for $x<0$ and $\rho>\alpha$, therefore data $(\rho,x)$ in this region of the plane
indicates the presence of a signal with unit probability. This intuition is clearly borne out in the above analysis since
\begin{equation}
p(\sig \given \rho, x) = \frac{p(\rho, x \given \sig) p(\sig)}
                              {p(\rho, x \given \sig) p(\sig) + 0} = 1
\end{equation}
for $\{ (\rho,x) \given x<0 \textrm{ and } \rho>\alpha\}$, compare this equation with Eq.~\eqref{bayes_prob}. The likelihood
ratio for these data points is infinite, reflecting complete certainty that the
data samples from this region are signals. 

\subsubsection{Simulated compact binary search}

For the purpose of simulating a real-life search we use data from LIGO's fourth science run, February
24--March 24, 2005. The data was collected by three detectors: the H1 and H2
co-located detectors in Hanford, WA, and the L1 detector in Livingston, LA.

The search targets three types of binaries: neutron star--neutron star
(BNS), neutron star--black hole (NSBH) and black hole--black hole (BBH). To model signals from these
systems, we use non-spinning, post-Newtonian
waveforms~\cite{Blanchet:1996pi,Droz:1999qx,Blanchet:2002av,Buonanno:2006ui,Boyle:2007ft,Hannam:2007ik,pan:024014,Boyle:2009dg,thorne.k:1987,SathyaDhurandhar:1991,Owen:1998dk} that are
Newtonian order in amplitude and second order in phase, calculated using the
stationary phase approximation~\cite{Droz:1999qx,
thorne.k:1987,SathyaDhurandhar:1991} with the upper cut-off frequency set by
the Schwarzschild innermost stable circular orbit. We generate three sets of simulated signals, one for each type of binary. The neutron star masses are chosen randomly in the
range 1--3~\Msun, while the black hole masses are restricted such that the total binary mass is between
2--35~\Msun. The maximum allowed distance for the source systems is
set to 20~Mpc for BNS, 25~Mpc for NSBH and 60~Mpc for BBH\@. These distances roughly correspond to the sensitivity range of the detectors in this science run. All other parameters, including the location of the source on the sky, are randomly sampled. The simulated signals are distributed uniformly in distance. In order to represent realistic astrophysical population with probability density function scaling as distance squared, the simulated signals are appropriately re-weighted and are counted according to their weights. The simulated signals from each set are injected into non-overlapping 2048-second blocks of data and analyzed independently. 

Analysis of the data is performed using the low-mass CBC 
pipeline~\cite{LIGOS3S4all, LIGOS3S4Tuning, Collaboration:2009tt,
Abbott:2009qj, S5LowMassLV}. It consists of several stages. First, the data
recorded by each interferometer is match-filtered with the bank of
non-spinning, post-Newtonian template waveforms covering all possible binary
mass combinations with total mass in the range 2--35~\Msun. The template
waveforms come from the same family as the simulated signal waveforms previously described. When the SNR time series for a particular template crosses the
threshold of 5.5, a single-interferometer trigger is recorded. This
trigger is then subjected to waveform consistency tests, followed by consistency testing with triggers from the other
interferometers. To be promoted to a gravitational-wave candidate, a signal is
required to produce triggers with similar mass parameters in at least two interferometers within a very
short time window (set by the light travel time between the detectors). The surviving coincident triggers are ranked
according to the combined effective \ac{SNR} statistic given by
\begin{equation}
\label{eq:combinedstat}
\rho_{\rm c}^2=\sum_{i=1}^N \rho_{{\rm eff}, i}^2\,,
\end{equation}
where the sum is taken over the triggers from different detectors that were
identified to be in coincidence and the phenomenologically constructed
\emph{effective \ac{SNR}} for a trigger is defined as
\begin{equation}
\label{eqn:effsnr}
\rho_{\rm eff}^2 = \frac{\rho^2}
{\sqrt{\left(\frac{\chi^2}{2p-2}\right)\left(1+\frac{\rho^2}{r}\right)}}\,,
\end{equation}
where $\rho$ is the \ac{SNR}, the phenomenological denominator factor $r =
250$, and $p$ is the number of bins used in the $\chi^2$ test, which is a
measure of how much the signal in the data matches the template~\cite{Allen:2004}. In the
denominator of Eq.~\eqref{eqn:effsnr}, $\chi^2$ is normalized by $2p -
2$, the number of degrees of freedom for this test.

All steps in the analysis beyond calculation of the \ac{SNR}, $\rho$, are
designed to remove non-Gaussian noise artifacts. Experience has shown that if
properly tuned, these extra steps significantly reduce the number of false
alarms~\cite{LIGOS3S4Tuning}. Yet typically, the resulting output of the
analysis is still not completely free of instrumental artifacts. Triggers that survived the pipeline's initial tests include unsuppressed noise artifacts. The general formalism developed in Section~\ref{GeneralDerivation} can be applied to further classify these triggers with the aim of optimally separating signals from the noise artifacts. Each trigger is characterized by a vector of parameters which, in addition to the combined effective \ac{SNR}, $\rho_{\rm c}$, may include the chirp mass, $\mc$, difference in the time of arrivals at different detectors etc. Such information as which detectors detected the signal and what was the data quality at the time of detection can be also folded in as a discrete trigger parameter.  For such parametrized data, the probability distributions in the presence and absence of a signal can be
estimated via direct Monte Carlo simulations. These distributions, if estimated
correctly, include a non-Gaussian component. The triggers are ranked
by their likelihood ratios, Eq.~\eqref{eq:likelihoodratio}, which results
in the optimized search in the parameter space of triggers.

Extra efficiency gained by additional processing of the triggers depends
strongly on the extent to which the non-Gaussian features of the background
noise are reflected in the distribution of the trigger parameters. In the
context of the search for gravitational waves from compact binary coalescence
in LIGO data, the chirp mass of a trigger's template waveform is one of the parameters that exhibits a
non-trivial background distribution. For a given $\mc$, the number of
background triggers falls off with increasing combined effective \ac{SNR}, $\rho_{\rm c}$, of the
trigger. The rate of falloff is slower for templates with
higher chirp mass, reflecting the fact that non-Gaussian noise artifacts are
more likely to generate a trigger for templates with smaller bandwidth. Another important piece of information about a trigger is the number and type of detectors that produced it. Generally, detectors differ by their sensitivities and level of noise. In the case we are concern with, two detectors, H1 and L1, have comparable sensitivities which are factor of two higher than the sensitivity of the smaller H2 detector. This configuration implies that the signals within the sensitivity range of the H2 detector are likely to be detected in all three instruments forming a set of triple triggers, H1H2L1. The signals beyond the reach of the H2 detector can only be detected in two instruments forming a set of double triggers, H1L1. Detection of a true signal by another two detector combinations, H1H2 and H2L1, is very unlikely, therefore such triggers  are discarded in the search. The number density of astrophysical sources grows as distance squared. As consequence, it is more likely that a gravitational-wave signal is detected as an H1L1 double trigger. On the other hand, background of H1H2L1 triggers is much cleaner due to the fact that instrumental artifacts are less likely to occur in all three detectors simultaneously. These competing factors should be included in the ranking of the candidate events in order to optimize probability of detection.

It is natural to expect that inclusion of such information about the triggers in the ranking, in
addition to the combined effective \ac{SNR}, should help distinguishing signals from 
noise artifacts. The first step is to estimate distribution of trigger parameters for signals and background. For background estimation, we use the time-shifted data---the standard technique employed in the searches for
transient gravitational-wave signals in LIGO data~\cite{LIGOS3S4all,
LIGOS3S4Tuning, Collaboration:2009tt, Abbott:2009qj, S5LowMassLV}. We perform
200 time shifts of the data recorded by L1 with respect to the data
taken by the H1 and H2 detectors. The time lags are multiples of 5 seconds. 

Analysis of time-shifted data provides us with a sample of the background
distribution of the combined effective \ac{SNR}s for H1L1 and H1H2L1 triggers with various chirp masses. We
find that all triggers can be subdivided into three chirp mass bins: $0.87
\leq \mathcal{M}_\mathrm{c}/\Msun < 3.48$, $3.48 \leq
\mathcal{M}_\mathrm{c}/\Msun < 7.4$, and $7.4 \leq \mathcal{M}_\mathrm{c}/\Msun
< 15.24$. These correspond to equal mass binaries with total masses of
2--8$~\Msun$, 8--17$~\Msun$ and 17--35$~\Msun$. These same 
bins were used in the analyses of the data from LIGO's S5 and Virgo's VSR1 science
runs~\cite{Collaboration:2009tt, Abbott:2009qj, S5LowMassLV}. Within each bin,
the background distributions depend weakly on chirp mass, thus there is no need
for finer resolution. At the same time, the distributions of the combined effective \ac{SNR} in
different bins show progressively longer tails with increasing chirp mass. 

The distribution of triggers for gravitational-wave signals is simulated by injecting model waveforms into the data and analyzing them with the pipeline. This is done independently for each source
type: BNS, NSBH and BBH\@.

Following the prescription for optimal ranking outlined in
Section~\ref{GeneralDerivation}, we treat each trigger as a vector of data $\vec{c} = (\rho_{\rm c}, \alpha, {\rm m})$, where $\alpha$ denotes the type of the trigger, double H1L1 or triple H1H2L1, and $\rm m$ is a discrete index labeling the chirp mass bins. We construct the likelihood-ratio ranking, $\Lambda(\rho_{\rm c}, \alpha,{\rm m} \given S_j)$ for each binary type, where $S_j$ stands for BNS, NSBH or BBH\@. Note that the likelihood ratio has strong dependence on the binary type, $S_j$. To simplify calculations, we approximate the likelihood ratio by 
\begin{equation}
\label{LikelihoodApproximation}
\Lambda(\rho_{\rm c}, \alpha, {\rm m} \given S_j) \approx \frac{n^j_{\textrm{inj}}(\rho_{\rm c}, \alpha, {\rm m})}{n_{\textrm{slide}}(\rho_{\rm c}, \alpha, {\rm m})}\,,
\end{equation}
where $n^j_{inj}(\rho_{\rm c}, \alpha, {\rm m})$ is the fraction of injected signals of $S_j$ type that
produce a trigger of type $\alpha$ with $\rho'_{\rm c} \geq \rho_{\rm c}$ in the chirp mass
bin $\rm m$, and $n_{slide}(\rho_{\rm eff}, \alpha, {\rm m})$ is the fraction
of time shifts of the data that produce a trigger of type $\alpha$ with $\rho'_{\rm c} \geq
\rho_{\rm c}$ in the same chirp mass bin. This approximation is equivalent to
using cumulative probability distributions instead of probability densities. It is expected to be reasonably good for the tails of probability distributions
that fall off as a power law or faster. The case we consider here falls in this
category. 

%\graphicspath{{likelihood_plots/}}
\begin{figure}
\includegraphics[width=3in]{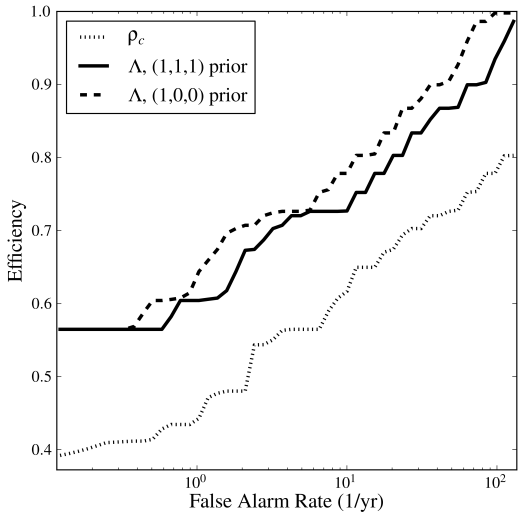}
\caption{Efficiency in detecting BNS signals versus false-alarm rate computed
for various rankings. The solid curve corresponds to the likelihood-ratio ranking, $\Lambda$, with uniform prior $p_s(S_j)=(1,1,1)$. The dashed curve is the likelihood-ratio ranking, $\Lambda$, with the prior $p_s(S_j)=(1,0,0)$, singling out BNS binaries for detection. The dotted curve represents the standard search with the combined effective \ac{SNR} ranking, $\rho_{\rm eff}$.}
\label{fig:bns_LVFAR}
\end{figure}

%\graphicspath{{likelihood_plots/}}
\begin{figure}
\includegraphics[width=3in]{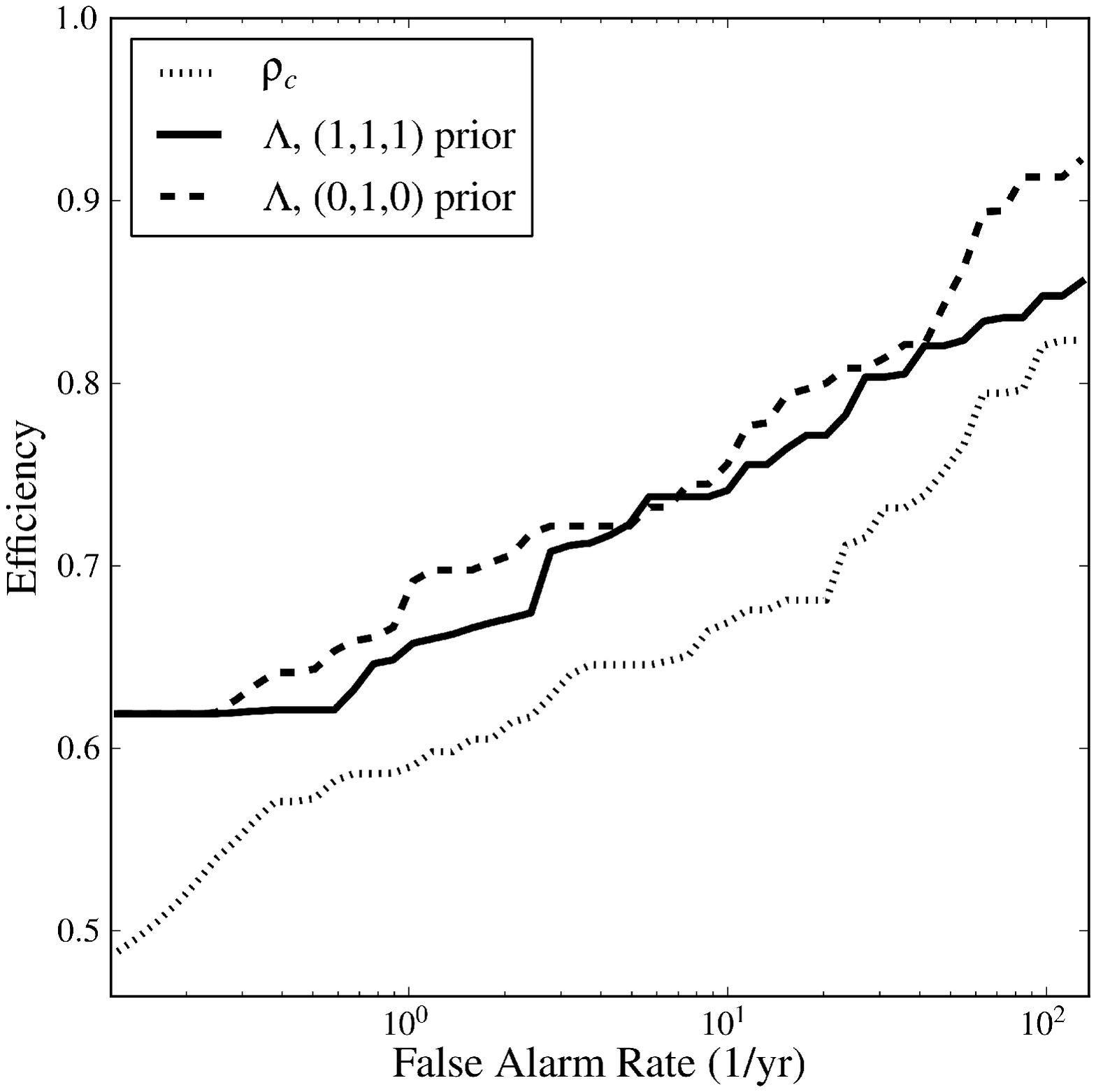}
\caption{Efficiency in detecting NSBH signals versus false alarm rate computed
for various rankings. The solid curve corresponds to the likelihood-ratio ranking, $\Lambda$, with uniform prior $p_s(S_j)=(1,1,1)$. The dashed curve is the likelihood-ratio ranking, $\Lambda$, with the prior $p_s(S_j)=(0,1,0)$, singling out NSBH binaries for detection. The dotted curve represents the standard search with the combined effective \ac{SNR} ranking, $\rho_{\rm eff}$.}
\label{fig:nsbh_LVFAR}
\end{figure}

%\graphicspath{{likelihood_plots/}}
\begin{figure}
\includegraphics[width=3in]{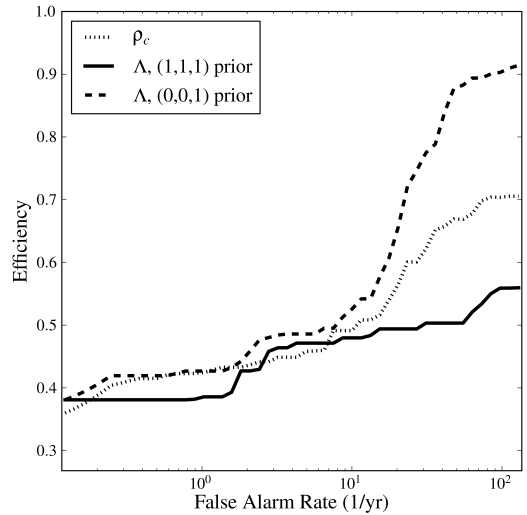}
\caption{Efficiency in detecting BBH signals versus false alarm rate computed
for various rankings. The solid curve corresponds to the likelihood-ratio ranking, $\Lambda$, with uniform prior $p_s(S_j)=(1,1,1)$. The dashed curve is the likelihood-ratio ranking, $\Lambda$, with the prior $p_s(S_j)=(0,0,1)$, singling out BBH binaries for detection. The dotted curve represents the standard search with the combined effective \ac{SNR} ranking, $\rho_{\rm eff}$. }
\label{fig:bbh_LVFAR}
\end{figure}

We apply the new ranking statistic given by Eq.~\eqref{LikelihoodApproximation}
to all triggers: background and signals. Each trigger has three likelihood ratios, one for each binary type. We introduce a prior distribution for binary types, $p_s(S_j)$. It can either encode our knowledge  about astrophysical   populations of binaries or relative ``importance'' of different types of binaries to the search. In what follows we consider four alternatives: $p_s(S_j) = (1,0,0)$, $p_s(S_j) = (0,1,0)$, $p_s(S_j) = (0,0,1)$ and $p_s(S_j) = (1,1,1)$. The first three singles out one of the binary types, whereas the last one treats all binaries on equal footing. Finally, the ranking statistic is defined as
 \begin{equation}
\label{Maxlikelihood}
\Lambda(\rho_{\rm c}, \alpha, {\rm m} )  = \max_{S_j} \Lambda(\rho_{\rm c}, \alpha, {\rm m} \given S_j)p_s(S_j)\,.
\end{equation}
The four alternative choices for $p_s(S_j)$ define four searches. For example, $p_s(S_j) = (1,0,0)$ corresponds to the search targeting only gravitational-wave signal from BNS coalescence. Similarly, $p_s(S_j) = (0,1,0)$ and $p_s(S_j) = (0,0,1)$ define the searches for gravitational waves from NSBH and BBH\@. The uniform prior, $p_s(S_j) = (1,1,1)$, allows one to detect all signals without giving priority to one type over the others. In each of the searches, the likelihood-ratio ranking, Eq.~(\ref{Maxlikelihood}), re-weights triggers giving higher priority to those that are likely to be the targeted signal as oppose to noise.  

In order to assess the improvement attained by the new ranking, we compute efficiency in recovering simulated signals from the data
as a function of the rate of false alarms. For a given rate of false alarms we find the corresponding value of the ranking and define efficiency as ratio of injected signals ranked above this value to the total number of signals that passed initial cuts of the analysis pipeline. This is equivalent to computing the
standard receiver-operator curve $P_1(P_0)$ defined by
Eqs.~\eqref{detprob}--\eqref{faprob}. The Efficiency curves are computed for BNS, NSBH and BBH binaries. In each case we evaluate efficiency of both likelihood-ratio rankings, the one that targets only that type of binary and the one that applies the uniform prior, $p_s(S_j)=(1,1,1)$. We compare the resulting curves to the efficiency curve for the standard
analysis pipeline that uses the combined effective \ac{SNR}, $\rho_{\rm c}$ as the ranking statistic. These curves  are shown in Figures~\ref{fig:bns_LVFAR}--\ref{fig:bbh_LVFAR}.

\begin{figure}
\includegraphics[width=3in]{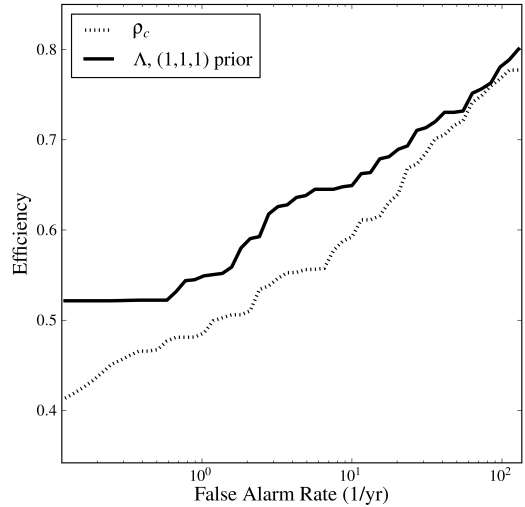}
\caption{Efficiency in detecting signals from any binary (BNS, NSBH or BBH) versus the false alarm rate computed
for various rankings. The solid curve corresponds to the likelihood-ratio ranking, $\Lambda$, with uniform prior $p_s(S_j)=(1,1,1)$. The dotted curve represents the standard search with the combined effective \ac{SNR} ranking, $\rho_{\rm eff}$.}
\label{fig:combined_LVFAR}
\end{figure}

They reveal that the searches targeting single type of binary, represented by the dashed curves, are more sensitive than the uniform search, the solid curve. This is expected, because narrowing down the space of signals typically allows one to discard the triggers that mismatch the signal's parameters reducing the rate of false alarms without loss of efficiency in recovering these signals. For instance, the search targeting BNS only signals discards all triggers from the high chirp mass bins, $\rm m = 2,3$, without discarding the BNS signals. This reduces the rate of false alarms, although at the prices of missing possible gravitational-wave signals from other types of binaries, NSBH and BBH\@. Still, one could justify such search if it was known that NSBH and BBH binaries do not exist or are very rare. The uniform search, despite being less sensitive to BNS signals, allows one to detect the signals from all kinds of binaries. Such search still gains in efficiency over the standard search, the dotted curve, for BNS and NSBH systems, Figures~\ref{fig:bns_LVFAR} and ~\ref{fig:nsbh_LVFAR}. At the same time, Figure~\ref{fig:bbh_LVFAR} shows that such search does worse in comparison to the standard search in detecting BBH signals. This is an unavoidable consequence of re-weighting of triggers by the likelihood ratio, Eq.~(\ref{Maxlikelihood}) based on their type and chirp mass. It ranks triggers from the lower chirp mass bins higher, because these triggers are less likely to be a noise artifact. This leads to some loss of sensitivity to BBH signals, but gains sensitivity to BNS and NSBH signals. The role of the likelihood ratio is to provide optimal re-weighting of triggers that results in the highest overall efficiency. In the case of the uniform search, it should provide increase in the total number of detected sources of all types. To demonstrate that this is indeed the case, we plotted the combined efficiency of the uniform search for BNS, NSBH and BBH signals and compared it to the efficiency of the standard search, Figure \ref{fig:combined_LVFAR}. 

%\graphicspath{{likelihood_plots/}}

\begin{figure}
\includegraphics[width=3in]{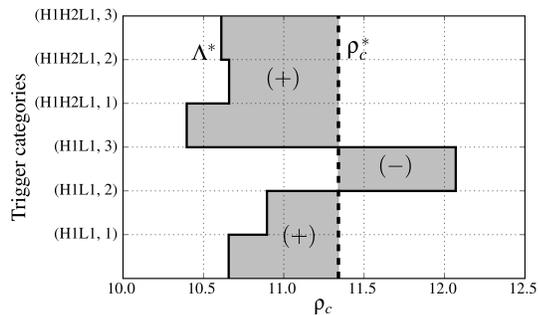}
\caption{The detection surfaces for the combined effective SNR, $\rho_{\rm_c}$, and the likelihood-ratio, $\Lambda$, rankings at the false alarm rate of 1.25 events per year. The y-axes labels different types and chirp mass bins of the triggers. The dashed line is the line of constant combined effective \ac{SNR}, $\rho_{\rm c}^* = 11.34$.  The solid line is the line of constant likelihood ratio, $\ln \Lambda^*(\rho_{\rm c}, \alpha, \rm m ) = 9.11$. The signal producing a trigger that falls to the right of the dashed/solid curve is considered to be detected in the search with combined effective \ac{SNR}/likelihood-ratio ranking. Those triggers that fall to the left are missed. The shaded region is the difference between the detection region for the likelihood-ratio and the combined effective \ac{SNR} rankings. The signals that produce trigger with parameters in the shaded regions labeled by $(+)$ are gained in the search equipped with likelihood ratio but missed by the search with the combined effective \ac{SNR} ranking. Those signals that produce a trigger in the shaded region labeled by $(-)$ are missed by the likelihood-ratio ranking but detected by the combined effective \ac{SNR} ranking.}
\label{fig:constant_lr_curve}
\end{figure}

The combined efficiency of the uniform search on Figure \ref{fig:combined_LVFAR} is higher than that of the standard search  because triggers are re-weighted by the likelihood ratio which properly accounts for the probability distributions of noise and signals. To gain further insight in this process we pick a particular point on the efficiency curve that corresponds to the rate of false alarms, x-axes, of $1.25$ events per year. We find the corresponding to this rate threshold for combined effective \ac{SNR} in the standard search to be $\rho_{\rm c}^* = 11.34$. Next, we find the corresponding threshold for logarithm of the likelihood ratio, $\ln \Lambda^*(\rho_{\rm c}, \alpha, \rm m ) = 9.11$. For each $(\alpha, \rm m)$ combination this value can be mapped to $\rho_{\rm c}$ which will be different for each type of trigger. Both $\rho_{\rm c}^* = 11.34$ and $\ln \Lambda^*(\rho_{\rm c}, \alpha, \rm m ) = 9.11$ define detection surfaces in $(\rho_{\rm c}, \alpha, \rm m )$ space of trigger parameters. We depicted them on Figure (\ref{fig:constant_lr_curve}). 

%\graphicspath{{likelihood_plots/}}

The signals falling to the right of $\rho_{\rm c}^* = 11.34$, the dashed line, are considered to be detected in the standard search. Similarly, the signals that happen to produce a trigger to the right of $\ln \Lambda^*(\rho_{\rm c}, \alpha, \rm m ) = 9.11$, the solid line, are considered to be detected in the uniform search with the likelihood-ratio ranking. The line of constant likelihood-ratio ranking sets different thresholds for combined effective snr, $\rho_{\rm c}$, of the triggers depending on their type. The threshold is higher than $\rho_{\rm c}^* = 11.34$ for the H1L1 triggers from the third chirp mass bin. The signals producing triggers in the shaded area in this bin, labeled by $(-)$, are missed in the uniform search but detected by the standard search. These signals are typically corresponds to BBH coalescence. The effect of this is visible on Figure (\ref{fig:bbh_LVFAR}), the solid curve is below the dotted curve at false alarm rate of $1.25$ events per year. On the other hand, the thresholds for other trigger types and chirp masses are lower than $\rho_{\rm c}^* = 11.34$. As a result, the signals producing triggers with parameters in the shaded regions labeled by $(+)$ are detected in the uniform search but missed by the standard one. The net gain from detecting these signals is positive, Figure \ref{fig:combined_LVFAR}. The process of optimization of the search in $(\rho_{\rm c}, \alpha, \rm m )$ parameter space can be thought of as deformation of the detection curve, $\rho_{\rm c}^* = 11.34$, with the aim of maximizing efficiency of the search. The deformations are constrained to those that do not change the rate of false alarms. The optimal  detection surface, as was shown in Section \ref{FrequentistApproach}, Eq.~(\ref{optcondition}), is the surface of constant likelihood ratio. This is the essence of likelihood ratio method. 

The power of the likelihood-ratio ranking depends strongly on the input data. For demonstration purpose, in the simulation we restricted our attention to a subset of trigger parameters,$(\rho_{\rm c}, \alpha, \rm m )$.  We expect that inclusion of other parameters such as difference in arrival times of the signal at different detectors, ratios of recovered amplitudes  etc, should drastically improve the search. We leave this to future work.

\section{Conclusion}

In this paper, we describe a general framework for designing optimal searches for
transient gravitational-wave signals in data with non-Gaussian background
noise. The principle quantity used in this method is the likelihood ratio, the ratio of the likelihood that the observed data contain signal to the likelihood that the data contain only noise. In Section~\ref{GeneralDerivation} we prove that the likelihood ratio leads to the optimal analysis of data, incorporating all available information. It is robust against increase of the data volume, effectively ignoring irrelevant information. We apply the general formalism to two typical problems that arise in searches for gravitational-wave signals in LIGO data. 

First, in Section~\ref{s:combining_disjoint_experiments} we show that when searching for gravitational-wave signals in the data
from different experiments or detector configurations, it is necessary as well
as sufficient to rank candidates by the ``local'' likelihood ratio given by Eq.~(\ref{timecatlikelihood}), which is calculated using estimated local probabilities. This provides overall optimality across the experiments. Candidate events from different
experiments can be compared directly in terms of their likelihood ratios. This
results in complete unification of the data analysis products. Another
significant feature of the unified analysis is that the candidate's
significance is independent of the duration of the experiments. Only the
detectors' sensitivities and level of background noise contribute to the
likelihood ratio of the candidates. The experiment's duration, on the other
hand, measures its contribution to the total probability of detecting a signal
(or efficiency) and the total probability of a false alarm. 

Second, in Section~\ref{s:combining_search_spaces} we aim to improve efficiency of the search for gravitational waves from compact binary coalescence by considering the issue of consistent accounting for non-Gaussian features of the noise in the analysis.  We suggest a practical solution to this problem. Estimate the probability distributions of parameters of the candidate events (e.g.
\ac{SNR} and the chirp mass of the template waveform, type of trigger etc) in the presence and absence
of a signal in the data. Construct the likelihood ratio that includes
non-Gaussian features and use it to re-rank candidate events.
Non-trivial information contained in the probability distributions of
candidate's parameters allows for a more optimal evaluation of their
significance. Indeed, as we demonstrate in the simulation, inclusion of the chirp mass and the type of trigger in the
likelihood-ratio ranking results in a significant increase of efficiency in
detecting signals from coalescing binaries. 

We would like to stress that the approach described in this paper is quite generic and can be applied to wide range of problems in analysis of data with non-Gaussian background. Its main advantage is consistent account of statistical information contained in the data. It provides a unified measure, in the form of the likelihood ratio, of the information relevant to detection of the signal in any type of data. This allows one to combine data of very different kind, such as the type of experiment, a type of trigger, its discrete and continuous parameters etc, into the single optimized analysis.

\acknowledgments
Authors would like to thank Ilya Mandel and Jolien Creighton for many fruitful discussions and helpful suggestions. This work has been supported by NSF grants PHY-0600953 and PHY-0923409. DK was supported from the Max Planck Gesellschaft. LP and RV was supported by LIGO laboratory. LIGO was constructed by the California Institute of Technology and Massachusetts Institute of Technology with funding from the National Science Foundation and operates under cooperative agreement PHY-0757058.

\bibliography{../../bibtex/iulpapers}

\end{document}